\documentclass[letterpaper, 11pt, published, accepted=2023-04-09]{quantumarticle}
\pdfoutput=1

\usepackage[utf8]{inputenc}
\usepackage[english]{babel}
\usepackage[T1]{fontenc}

\usepackage{amsmath}
\usepackage{amssymb}
\usepackage{amsthm}
\usepackage{physics}
\usepackage{mathtools}
\usepackage{subcaption}
\usepackage{booktabs}

\usepackage{hyperref}
\usepackage[numbers,sort&compress]{natbib}

\newcommand{\code}{\ensuremath{\mathcal C}}
\newcommand{\const}[1]{\boldsymbol{c_{#1}}}

\newcommand{\edge}[2]{\ensuremath{\qty{#1, #2}}}

\newcommand{\generators}[1]{\ensuremath{g(#1)}}
\newcommand{\group}[1]{\ensuremath{\mathcal{#1}}}

\newcommand{\lxor}{\ensuremath{\oplus}}
\newcommand{\neighbors}{\ensuremath{\eta}}
\newcommand{\pauligroup}{\group{P}}
\newcommand{\stabgen}{\generators{\stabgroup}}
\newcommand{\stabgroup}{\group{S}}
\newcommand{\vari}[1]{\boldsymbol{v_{#1}}}

\DeclarePairedDelimiterX\set[1]\lbrace\rbrace{\def\given{\;\delimsize\vert\;}#1}

\theoremstyle{definition}
\newtheorem{definition}{Definition}

\begin{document}

\title{Finite-rate sparse quantum codes aplenty}

\author{Maxime Tremblay}

\author{Guillaume Duclos-Cianci}

\author{Stefanos Kourtis}
\email{stefanos.kourtis@usherbrooke.ca}

\affiliation{Département de physique \& Institut quantique, Université de Sherbrooke, Sherbrooke, Québec, Canada, J1K 2R1}

\maketitle

\begin{abstract}
	We introduce a methodology for generating random multi-qubit stabilizer codes based on solving a constraint satisfaction problem (CSP) on random bipartite graphs. This framework allows us to enforce stabilizer commutation, $X/Z$ balancing, finite rate, sparsity, and maximum-degree constraints simultaneously in a CSP that we can then solve numerically.
	Using a state-of-the-art CSP solver,
	we obtain convincing evidence for the existence of a satisfiability threshold.
	Furthermore, the extent of the satisfiable phase increases with the number of qubits.
	In that phase, finding sparse codes becomes an easy problem. Moreover, we observe that the sparse codes found in the satisfiable phase practically achieve the channel capacity for erasure noise. Our results show that intermediate-size finite-rate sparse quantum codes are easy to find, while also demonstrating a flexible methodology for generating good codes with custom properties. We therefore establish a complete and customizable pipeline for random quantum code discovery.
\end{abstract}

\section{Introduction}

Quantum error-correcting codes are an essential prerequisite of  reliable quantum computing.
While codes like surface codes~\cite{bravyi_quantum_1998,fowler_surface_2012,stephens_fault-tolerant_2014}
and color codes~\cite{landahl_fault-tolerant_2011, kubica_three-dimensional_2018} are sufficient to encode and protect a small constant number of qubits,
their performance degrades rapidly when performing
computations with more qubits~\cite{delfosse_tradeoffs_2013, bravyi_tradeoffs_2010}.
While quantum error-correcting codes with asymptotically good performance
have been known to exist for more than 20 years~\cite{calderbank_good_1996, ashikhmin_asymptotically_2001}, these codes require measurement of operators whose weight increases with block size.
This is a major limitation to efficient implementation of measurement circuits for these codes.

More recently, Gottesman showed that finite-rate quantum low-density parity-check (LDPC)
codes can be used to build constant-overhead fault-tolerant
quantum computers~\cite{gottesman_fault-tolerant_2013}. Quantum
LDPC codes are stabilizer codes with bounded-weight stabilizers.
That is, they involve only bounded-weight measurements.

This seminal result together with the plethora of
asymptotically optimal classical LDPC code
constructions~\cite{gallager_low-density_1962, richardson_modern_2008}
led to a surge of research toward quantum LDPC codes inspired
by their classical counterparts.
This resulted in many quantum LDPC code constructions
such as hypergraph product codes~\cite{tillich_quantum_2014},
homological product codes~\cite{bravyi_homological_2014}
and fiber bundle codes~\cite{hastings_fiber_2020}.
Finally, more than 20 years after the introduction of the first
quantum error-correcting codes,
Panteleev and Kalachev~\cite{panteleev_asymptotically_2022}
introduced the first family of quantum LDPC codes
with optimal asymptotic performance.

These constructions are based on homological products
of classical or quantum error-correcting codes.
While there is some flexibility in the choice of the
input codes,
these methods are hard to adapt to precise connectivity limitations
of near to mid-term quantum devices.
Furthermore,
they are generally more relevant for a larger number of qubits.
For example,
recent numerical studies of hypergraph product
codes~\cite{grospellier_combining_2020, tremblay_constant-overhead_2021}
involve a few thousands to hundreds of thousands of qubits.
Other constructions such as building codes from
low-depth random circuits~\cite{gullans_quantum_2021, brown_short_2013},
have been proposed but,
to our knowledge,
no procedure generates the stabilizer operators directly
while considering arbitrary physical limitations.

In this work,
we introduce a methodology for the generation of an arbitrary number of stabilizer operators
directly for any given number of qubits.
We reformulate code generation as a constraint satisfaction problem (CSP) on random bipartite graphs of varying edge inclusion probability,
whose vertices correspond to qubits and stabilizers.
The constraints imposed correspond to desired code properties,
including, but not limited to,
stabilizer commutation,
$X/Z$ balancing,
finite rate,
sparsity, and maximum-degree bounds.
The resulting CSP is akin to paradigmatic NP-complete problems like random $k$-SAT and random 2-coloring of $k$-uniform hypergraphs.
Although these CSPs are hard to solve in the worst case,
they often display a transition between a hard phase,
where problem instances are hard to solve on average, and an easy phase,
within which typical instances can be solved in polynomial time,
as a function of some parameter~\cite{achlioptas_rigorous_2005, achlioptas_random_2006}.
This motivates us to ask whether our code generation CSP on random bipartite graphs also exhibits a transition to an easy phase as a function of some parameter,
which in the present case we choose to be the edge inclusion probability of the sampled graphs.

To define our problem instances,
we sample bipartite graphs at random with varying edge inclusion probability, define appropriate constraints on qubit and stabilizer vertices that ensure our desired code properties, and then solve the resulting instances using a state-of-the-art CSP solver.
We discover a satisfiability threshold at a critical edge inclusion probability that decreases with increasing number of qubits. We therefore show that the CSP has an easy phase, whose extent grows with increasing number of qubits and within which we readily find finite-rate stabilizer codes.
Furthermore,
we find that the resulting codes are much sparser than the initial
graphs. Going a step further, we show that we can guarantee code sparsity via maximum-degree constraints, at the expense of a computational overhead.
Finally, we demonstrate that the codes we obtain using our methodology achieve the erasure channel capacity. We therefore establish a complete and customizable pipeline for random quantum code discovery that can be geared towards near- to mid-term quantum processor layouts.

The rest of this paper is organized as follows.
In Section~\ref{sec:stabcodes},
we recall notions of quantum coding theory and stabilizer code construction.
In Section~\ref{sec:stabilizing_edge_coloring},
we introduce our methodology for the generation of stabilizer codes through solving a CSP on
random bipartite graphs.
In Section~\ref{sec:csp},
we present an explicit construction of the constraints for CSS codes.
Finally,
we present numerical results on code discovery and near-optimal error correction performance for the erasure channel in Section~\ref{sec:results}.

\section{Stabilizer codes}
\label{sec:stabcodes}

\begin{figure}
	\begin{center}
		\includegraphics{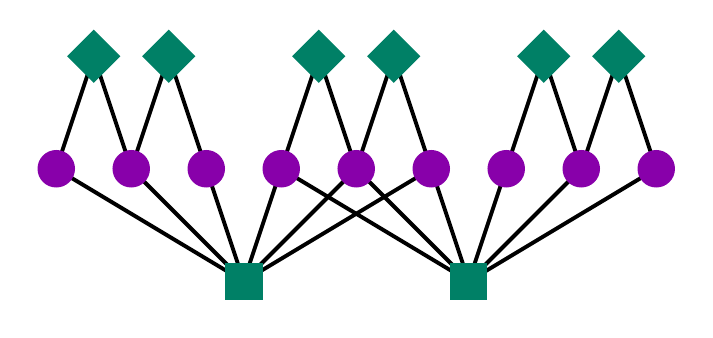}
	\end{center}
	\caption{
		The Tanner graph of the 9-qubit Shor code.
		Purple circles represent qubit vertices
		and green squares and diamonds represent respectively $X$ and $Z$
		stabilizer generators.
		The top right stabilizer vertex corresponds to the
		$Z_7 Z_8$ operator and the bottom left vertex
		corresponds to the $X_0 X_1 X_2 X_3 X_4 X_5$ operator.
	}
	\label{fig:tanner_graph}
\end{figure}

Given the $n$-qubit Pauli group $\pauligroup_n$,
a stabilizer group is a commuting subgroup of $\pauligroup_n$
that does not include the $-I$ operator.
A stabilizer group $\stabgroup$ defines the stabilizer code~\cite{gottesman_stabilizer_1997}
\begin{equation}
	\code(\stabgroup) = \set{
		\ket{\psi}
		\given
		S \ket{\psi} = \ket{\psi},
		\forall S \in \stabgroup
	}.
	\label{eq:stabcode}
\end{equation}
That is, a stabilizer code is the common $+1$ eigenspace of
the operators of a stabilizer group. We often define a stabilizer group using a set of generators
$\stabgen = \set{S_1, S_2, ..., S_m}$ with $S_i \in \stabgroup$.

A family of codes is a finite-rate
family if there exists a constant $c > 0$ such that $\frac{n - m}{n} > c$
for $n\to\infty$.
Finite-rate code families are crucial for achieving large-scale fault-tolerant
quantum computation since they allow encoding a constant
ratio of logical qubits to physical qubits with vanishing error rate.
This is in contrast with zero-rate code families such as surface codes, which can only encode
a small constant number of logical qubits
without significantly degrading error-correcting performance~\cite{bravyi_tradeoffs_2010}.

A common class of stabilizer codes are Calderbank-Shor-Steane (CSS)
codes~\cite{calderbank_good_1996,steane_simple_1996}.
These are codes for which there exists a set of generators such that
each of them is a product of only $I$ and $X$ or only $I$ and $Z$.
While the techniques introduced in this work are applicable to both CSS and non-CSS stabilizer codes alike,
in what follows we focus mainly on CSS codes as this simplifies the presentation of our methodology.
We discuss the case of non-CSS stabilizer codes in Sec.~\ref{sec:conclusion}.

Below we make use of the graphical representation of stabilizer codes based on Tanner graphs,
illustrated in Figure~\ref{fig:tanner_graph}.
A Tanner graph $T = (Q \cup \stabgen, E)$ is a bipartite graph that contains
edge $\edge{q}{S} \in E$ if and only if the stabilizer generator $S$ acts as $X$, $Y$ or $Z$ on
qubit $q$.
The degree of a vertex is its number of neighbors.
In error-correction terms,
the degree of a stabilizer corresponds to its weight and the degree
of a qubit corresponds to the number of stabilizers acting on it.

\section{Stabilizing edge coloring}
\label{sec:stabilizing_edge_coloring}

To introduce our algorithmic approach to principled search for finite-rate stabilizer codes,
we reformulate the task as a graph coloring problem.
Our strategy is to start with a bipartite graph $G = (Q \cup \stabgen, E_0)$,
called the support graph,
then search for an edge coloring $l : E_0 \to \qty{I, X, Y, Z}$ for which all stabilizers commute.
We call such a coloring a stabilizing edge coloring.
The result of this procedure is a Tanner graph $T = (Q \cup \stabgen, E)$
where the edge $e \in E$ iff $l(e) \neq I$ and hence $E \subseteq E_0$.
Deciding whether there exists
a stabilizing edge coloring for a given support graph $G$ is equivalent to determining whether there exists a valid stabilizer code whose Tanner graph is a subgraph of $G$ as defined above.

As formulated above, stabilizing edge coloring admits trivial solutions. For example, assigning the same color to all
edges is always a valid solution, but is of little interest for quantum error correction.
To avoid such trivial solutions, or other undesirable solutions,
we must introduce additional constraints.
We represent a constraint as a pair $(F, L)$ where $F \subseteq E_0$
and $L \subseteq \qty{I, X, Y, Z}^{|F |}$.
A coloring $l$ satisfies a constraint $(F, L)$ for
$F = \qty{e_1 , \dots, e_{|F|}}$ if $(l(e_1 ), . . . , l(e_{|F|})) \in L$.
We can now define the constrained stabilizing edge coloring problem we are interested in.

\begin{definition}[(Constrained) stabilizing edge coloring]
	\hfill\\
	\textbf{Given}\\
	a bipartite graph $G = (Q \cup \stabgen, E_0 )$ and a set of constraints
	$\mathcal{F} = \qty{(F_1 , L_1), \ldots, (F_{|\mathcal F|}, L_{|\mathcal F|})}$
	with $F_i \subseteq E_0$  and $L_i \subseteq \qty{I, X, Y, Z}^{|F_i|}$, \\
	\textbf{find}\\
	a stabilizing edge coloring $l : E_0 \to \qty{I, X, Y, Z}$
	\textbf{satisfying} $\mathcal{F}$.
	\label{def:stab_coloring}
\end{definition}

The problem is constrained whenever $\mathcal{F}$ is not empty.
Table~\ref{table:constraint} presents estimates of $|\mathcal F|$
for the instances we study numerically.
This defines a rich class of constraint satisfaction problems.
In the following section,
we provide concrete realisations of unconstrained and constrained stabilizing edge coloring
which we solve to generate non-trivial CSS codes.

\begin{figure}
	\begin{center}
		\includegraphics{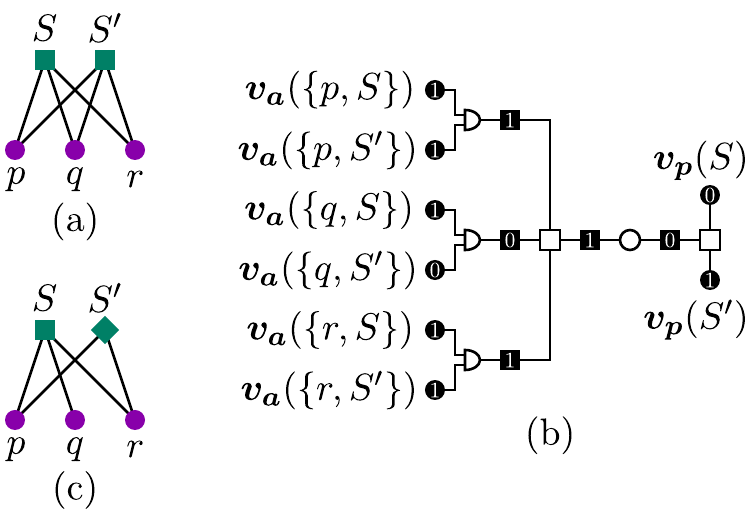}
	\end{center}
	\caption{
		Graphical representation of the commutation constraints for a pair of stabilizers
		sharing three qubits.
		(a)
		The support graph $G$.
		The circles and squares respectively represent qubit and stabilizers.
		(b)
		The boolean variables and constraints assuring the commutation.
		Filled vertices correspond to variables. In particular circles are either activator
		or Pauli variables while the squares are auxiliary variables.
		The circle constraints correspond to Equation~\eqref{eq:comm_constraint},
		the square constraints to Equations~\eqref{eq:stab_constraint} and \eqref{eq:even_constraint}
		and the half-circle constraints to Equation~\eqref{eq:both_active_constraint}.
		The numbers within variable vertices illustrate a valid assignment.
		(c)
		The resulting Tanner graph according to the variable assignment.
	}
	\label{fig:csp_commutation}
\end{figure}

\section{CSS codes from constraints on boolean variables}
\label{sec:csp}

For CSS codes,
we simplify stabilizing edge coloring by assigning Pauli values $l_p : \stabgen \to \qty{X, Z}$
to stabilizer vertices and boolean values $l_a : E_0 \to \qty{0, 1}$ to edges.
Then, an edge $e \in E_0$ has color $l(e) = I$ when $l_a(e) = 0$ and color $l_p(e)$ otherwise.
We say that an edge $e$ is active if $l_a(e) = 1$.
In this representation,
the active neighborhood of a stabilizer $S$ is the set of qubits for which $\{q, S\} \in E_0$ and $l_a(\{ q,S \}) = 1$.
Then,
two stabilizers with different Pauli values commute if the intersection
of their active neighborhoods has even cardinality.

To solve the stabilizing edge coloring problem numerically,
we represent the coloring functions $l_p$ and $l_a$ with boolean variables.
To each edge $e \in E_0$,
we assign a boolean variable $\vari{a}(e) \in \qty{0, 1}$ such that $l_a(e) = \vari{a}(e)$. To each stabilizer $S \in g(\mathcal{S})$, we assign a boolean variable $\vari{p}(S)$
such that $l_p(S) = X$ if $\vari{p}(S) = 1$ and $l(S) = Z$ otherwise.
We call $\vari{a}$ an activator variable and $\vari{p}$ a Pauli variable. We also introduce some auxiliary boolean variables. The auxiliary variable $\vari{s}(S, S') = 1$ when the action of $S$ and $S'$ is the same. The variable $\vari{e}(S, S') = 1$ when $S$ and $S'$ have an even overlap. Finally, the variable $\vari{b}(q, S, S') = 1$ when both edges $\set{q, S}$ and $\set{q, S'}$
are active.

We are now ready to introduce the boolean constraints that define the stabilizing edge coloring problem for CSS codes. In the rest of this section,
we represent a constraint as a boolean function $\const{}: \set{0, 1}^* \to \set{0, 1}$ acting
on a subset of the boolean variables $\vari{a}$, $\vari{p}$, $\vari{s}$, $\vari{e}$, and $\vari{b}$.
An assignment $\vb x \in \qty{0, 1}^*$ satisfies a constraint if $\const{}(\vb x) = 1$.
Boolean constraints allow us to represent the stabilizing edge coloring constraints
of Definition~\ref{def:stab_coloring}.

Stabilizers $S$ and $S'$ commute if they satisfy at least one of the following conditions:
they have the same non-trivial action ($X$ or $Z$) or,
as discussed previously,
they act non-trivially on an even number of common-neighbor qubits. We can thus write a commutation constraint $\const{c}(S, S')$ as
\begin{equation}
	\const{c}(S, S')
	=
	\vari{s}(S, S') \lor \vari{e}(S, S').
	\label{eq:comm_constraint}
\end{equation}
That the variable $\vari{s}(S, S')=1$ when the action of $S$ and $S'$ is the same is enforced by the constraint
\begin{equation}
	\const{s}(S, S')
	=
	\vari{s}(S, S') \lxor \vari{p}(S) \lxor \vari{p}(S').
	\label{eq:stab_constraint}
\end{equation}
Similarly, to ensure that $\vari{e}(S, S') = 1$ when $S$ and $S'$ have an even overlap, we add the constraint
\begin{equation}
	\const{e}(S, S')
	=
	\vari{e}(S, S')
	\lxor
	\qty[\bigoplus_{q \in \neighbors(S) \cap \neighbors(S')} \vari{b}(q, S, S')],
	\label{eq:even_constraint}
\end{equation}
where $\neighbors(v)$ is the set of neighbors of vertex $v$ in $G$.
Finally, to ensure that $\vari{b}(q, S, S') = 1$ when both edges $\set{q, S}$ and $\set{q, S'}$ are active, we add the constraint
\begin{align}
	\const{b}(q,S, S')
	=
	 & \lnot\vari{b}(q, S, S') \lxor \notag                \\
	 & [\vari{a}(\set{q, S}) \land \vari{a}(\set{q, S'})].
	\label{eq:both_active_constraint}
\end{align}

Equations~\eqref{eq:comm_constraint} to \eqref{eq:both_active_constraint}
define a set of constraints on activator, Pauli and auxiliary variables. We draw the variables and constraints for a small example in Figure~\ref{fig:csp_commutation}.
When simultaneously satisfied,
they ensure that the coloring function $l$ defines a stabilizing edge coloring.
Thus, any variable assignment for which all constraints evaluate to 1 yields a valid CSS code.
In Section~\ref{sec:cspsolver}, we present how we find such assignments.

\subsection{Extra constraints for good codes}

In this section,
we introduce extra constraints to restrict the search to better codes.
We represent these contraints using integer linear functions $\const{}: \mathbb Z^* \to \mathbb Z$
restricted to $D \subseteq \mathbb Z$.
That is,
an assignment $\vb x \in \mathbb Z^*$ satisfies constraint $\vb c$ if $\vb c(\vb x) \in D$.
We define these constraints by extending the boolean variables of the previous section to be integer variables.

We first add lower bounds on the number of stabilizers of each kind connected to a qubit.
For each edge $\set{q, S} \in E_0$,
we add a variable $\vari{X}(\set{q, S})$
with value $1$ if the edge is active and
the corresponding stabilizer is of the $X$ type.
This is enforced by the constraint
\begin{align}
	\const{X}(q,S)
	=
	\lnot \vari{X}(q, S) \lxor [\vari{a}(\set{q, S}) \land \vari{p}(S)].
	\label{eq:x_constraint}
\end{align}
Then,
we impose that the number of variables $\vari{X}(\set{q, S})$
with value 1 for $S \in \neighbors(q)$ is at least $\delta_q$.
That is,
we add the constraint
\begin{align}
	\const{X}(q) =
	\sum_{S \in \neighbors(q)} \vari{X}(q, S)
	\geq \delta_q
	\label{eq:x_bound_constraint}
\end{align}
for each qubit $q$.
We use similar constraints to lower bound the number of $Z$ stabilizer generators per qubit.

These constraints together with the commutation constraints are the first
set of constraints we numerically study in the following section.
Subsequently, we add more constraints to search for codes with improved
decoding performances.

We impose to each stabilizer $S \in \stabgen$ that the number of variables
$\vari{a}(\set{q, S})$ with value 1 for $q \in \neighbors(S)$ is at least $\delta_s$.
Then, to keep the codes sparse,
we impose that at most $\Delta_s$ edges per stabilizer are active.
These are both enforced by constraints akin to Equation~\eqref{eq:x_bound_constraint}.

Finally, we impose that the number of stabilizers of each kind is balanced
by enforcing that
\begin{equation}
	\sum_{S \in \stabgen} \vari{p}(S) = \left\lfloor\frac{|\stabgen|}{2}\right\rfloor.
\end{equation}

\begin{figure}[t]
	\begin{center}
		\includegraphics{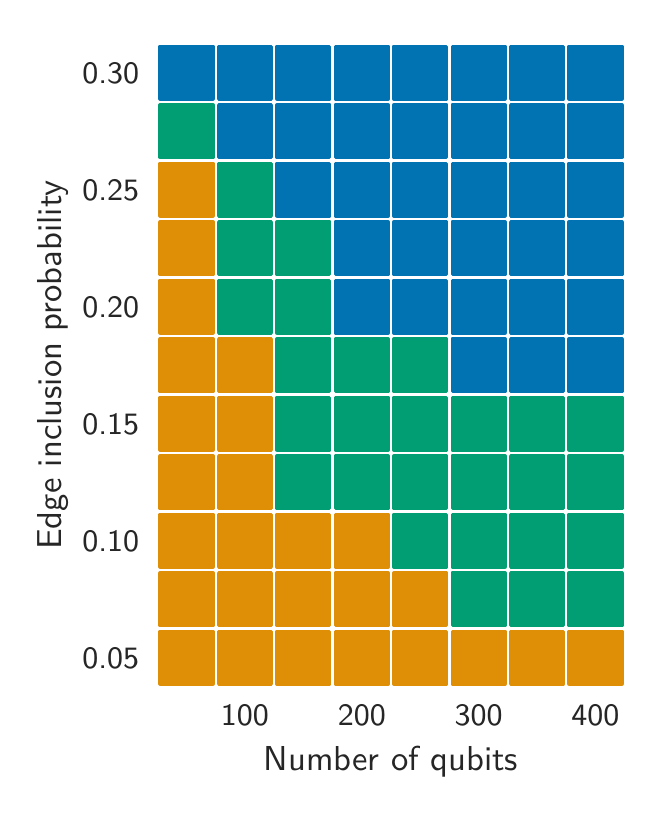}
		\label{fig:sat_phase_only_qubit_deg}
	\end{center}
	\caption{
		Satisfiability phase diagram for the commutation and minimum qubit
		degree constraints ($\delta_q = 3$).
		For each pixel we generate 100 support graphs, then solve the corresponding stabilizing edge coloring instance for each graph.
		For each support graph, we run the CSP solver on four CPU cores running at 2.4 GHz for up to four hours.
		A green pixel indicates a combination of edge inclusion probability and number of qubits for which the solver is able to solve less than 10\% of the instances within the allocated timeout.
		In the rest of the phase diagram, where more than 90\% of the instances are solved for each combination of parameters,
		a blue pixel indicates more satisfiable than unsatisfiable instances, whereas
		an orange pixel indicates the opposite: more unsatisfiable than satisfiable instances.
	}
	\label{fig:sat_phase_qubit_deg}
\end{figure}

\subsection{Constraint satisfaction problem solver}
\label{sec:cspsolver}

To obtain CSS codes from the aforementioned constraints,
we use the \textit{OR-Tools} library~\cite{noauthor_or-tools_2022}.
We decompose the constraints of Equations~\eqref{eq:comm_constraint}, \eqref{eq:both_active_constraint},
and \eqref{eq:x_constraint} into a constant number of OR constraints
and we leave those of Equations~\eqref{eq:stab_constraint} and \eqref{eq:even_constraint} as XOR constraints.
Both OR and XOR constraints,
as well as the linear constraints used to enforce the minimum and maximum degree and balancing,
are native to the library.
We provide our implementation in an online repository~\cite{noauthor_stabilizer_2022}.

\section{Results}
\label{sec:results}

\subsection{Phase transition}

To search for codes with $n$ qubits and $m$ stabilizer generators,
we start by building random support graphs with the corresponding numbers of vertices
using the Erdős–Rényi model~\cite{paul_erdos_random_1959}.
That is,
we sample the graphs with an edge inclusion probability of $\gamma$.
We use $G_{n,m,\gamma}$ to denote the corresponding random bipartite graph generator.
Our goal is to find sparse codes for given $n,m$, and $\gamma$,
or obtain a proof that such codes are statistically unlikely.

One can see that
if $E \subseteq E'$ and $G = (V, E) \in \mathcal P$, where $\mathcal{P}$ denotes the existence of at least one stabilizing edge coloring, then
for any graph $G' = (V, E')$, we have $G' \in \mathcal P$.
Thus,
\begin{equation}
	\Pr[G_{n,m,\gamma} \in \mathcal P]
	\leq
	\Pr[G_{n,m,\gamma'} \in \mathcal P]
\end{equation}
when $\gamma \leq \gamma'$ and there exists a threshold function
$\gamma^*(n)$ such that~\cite{bollobas_threshold_1987}
\begin{equation}
	\lim_{n\to \infty}\Pr[G_{n,m,\gamma} \in \mathcal P]
	=
	\begin{cases}
		0 & \gamma(n)/\gamma^*(n) \to 0,      \\
		1 & \gamma(n)/\gamma^*(n) \to \infty.
	\end{cases}
\end{equation}
Our numerical results, discussed below, suggest this threshold is a non-increasing function of the number of qubits.

We start by imposing only the commutation constraints and a minimum qubit degree.
That is, we impose $\delta_q = 3$, $\delta_s = 0$ and $\Delta_s = \infty$ for
graphs with fixed ratio $m/n = 9/10$, yielding codes with rate $1/10$.
If no solution was found and no proof of unsatisfiability was produced within a timeout,
we label the instance as unknown.
We note that most instances are solved in a fraction of the timeout,
except close to the threshold.
In Table~\ref{table:constraint} we give estimates for the number and weight (in number of variables) of each type of constraint in the CSS stabilizing edge coloring for the Erdős–Rényi model.

\begin{table}
	\begin{center}
		\begin{tabular}{cccc}
			\toprule
			                  & Type   & Weight      & Occurrence     \\
			\midrule
			Stab. edge color. & OR     & 3           & $nm^2\gamma^2$ \\
			                  & OR     & 2           & $m^2$          \\
			                  & XOR    & $n\gamma^2$ & $m^2$          \\
			                  & XOR    & 3           & $m^2$          \\
			$X/Z$ qubit deg.  & Linear & $m\gamma$   & $n$            \\
			                  & OR     & 3           & $nm\gamma$     \\
			Stab. deg.        & Linear & $n\gamma$   & $m$            \\
			Balancing         & Linear & $m$         & 1              \\
			\bottomrule
		\end{tabular}
	\end{center}
	\caption{Expected constraint occurrences and weights.
		The data is obtained considering there are $\mathcal O(m^2)$ pairs of stabilizers
		each expected to share $\mathcal O(n\gamma^2)$ common qubits.
	}
	\label{table:constraint}
\end{table}

\begin{figure}[t]
	\begin{center}
		\includegraphics{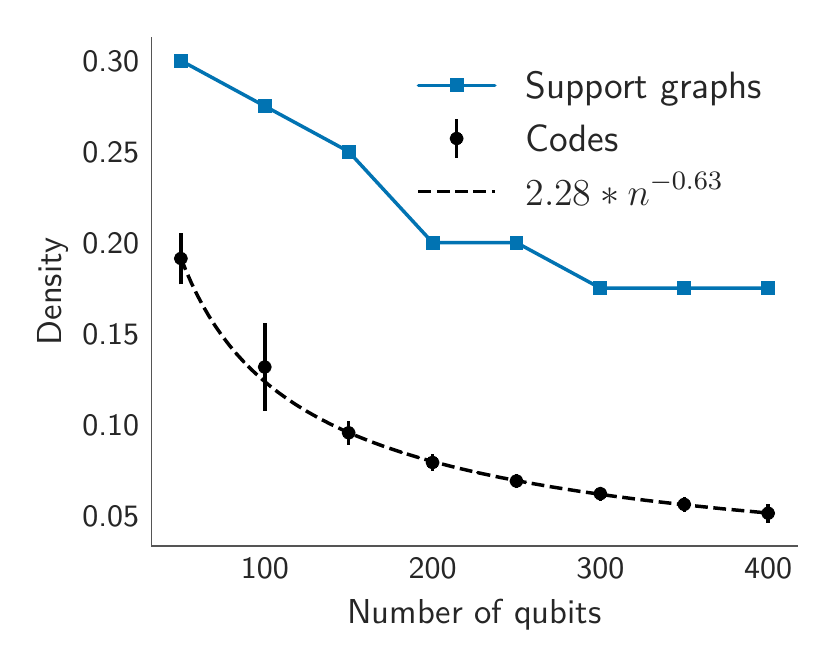}
	\end{center}
	\caption{
		Densities of the resulting codes for commutation and minimum qubit degree
		constraints.
		For the support graphs,
		we plot the minimum edge inclusion probability in the satisfiable region.
		For the codes,
		we plot the average density over all solutions with the same number of qubits found in the satisfiable phase of Figure~\ref{fig:sat_phase_qubit_deg}.
	}
	\label{fig:degrees}
\end{figure}

Figure~\ref{fig:sat_phase_qubit_deg} illustrates that the threshold $\gamma^*$ is decreasing with the number of qubits.
While we cannot definitively claim that the threshold does not start increasing for larger block sizes, we expect that finite-size effects are insignificant for larger numbers of qubits. This implies that $\gamma^*$ either decreases monotonically or it plateaus to a value of at most $15\%$.
The unknown region demarcates the parameter regime where typical instances of the CSP become hard to solve and our calculations time out.
Crucially, Figure~\ref{fig:degrees} shows that the densities $\frac{E}{mn}$ of the Tanner graphs of the codes go to
zero as the number of qubits increase and are much lower than the edge inclusion probabilites of the support graphs.
We observe similar results for different values of $m/n$.

\begin{figure}
	\begin{center}
		\includegraphics{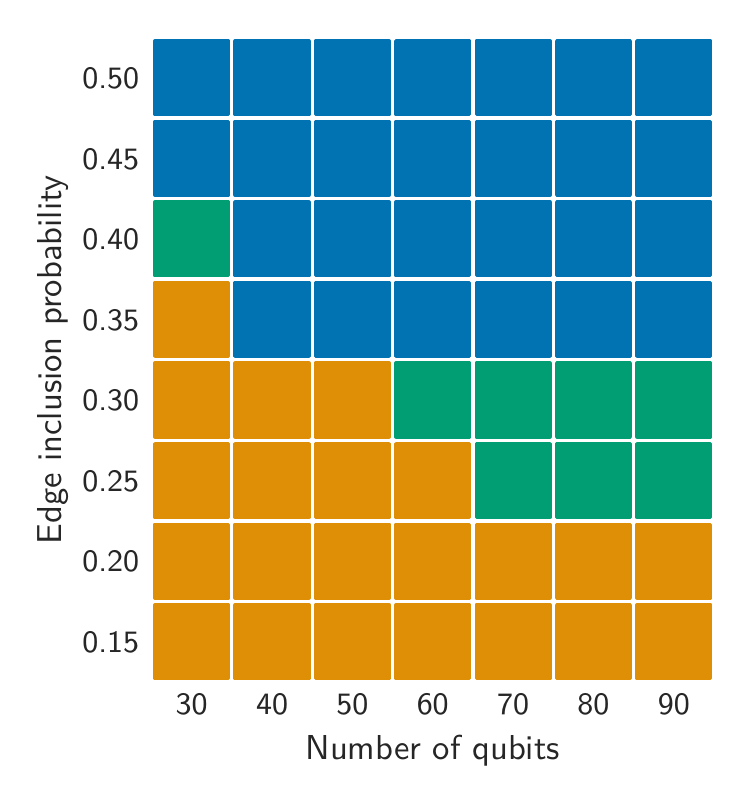}
	\end{center}
	\caption{
		Satisfiability phase diagram with qubit degree ($\delta_q = 3$), stabilizer degree ($\delta_s = 6, \Delta_s=20$), and balancing constraints.
		All parameters and details are the same as in Figure~\ref{fig:sat_phase_qubit_deg}.
	}
	\label{fig:sat_phase}
\end{figure}

We now restrict the problem further to the regime of quantum LDPC codes.
That is,
we impose $\delta_q = 3$, $\delta_s = 6$ and $\Delta_s = 20$ together
with the stabilizer balancing constraint.
Figure~\ref{fig:sat_phase} indicates that the threshold
function is non-increasing when increasing the number of qubits.
We thus once again find a satisfiable phase that expands with increasing number of qubits and within which it is statistically likely that quantum LDPC codes exist and are easy to find.

We also investigated the impact of different degree bounds.
However, due to significant increasing in required computing ressources,
we limited ourself to much fewer samples,
but we still observed similar results
(see Appendix~\ref{sec:extra}).
Also note that in practice,
it is not necessary to perform a large sampling
if one is interested only in finding a single good code.
Finally, although a maximum degree of 20 could be too large for some pratical applications,
the codes used in the decoding experiment of the following section have average degrees
between 8.4 and 12.7.

\subsection{Decoding experiment}

We showed numerical evidence that even if it is generally hard to find commuting sets
of random stabilizer operators,
it is relatively easy to generate such a set from a random bipartite graph
as long as the edge inclusion probability of the graph is above some threshold function.
However,
this result implies nothing about the decoding performance of the code
generated this way.

In this section, we probe the performance of the codes we find, using the erasure channel as suggested
in~\cite{delfosse_linear-time_2020, gullans_quantum_2021}.
The erasure channel is particularly useful since there exists a maximum-likelihood decoder
that runs in polynomial time.
In contrast, we do not yet have a good universal decoder for the depolarizing channel.

The single-qubit erasure channel,
\begin{equation}
	\mathcal{E}_p(\rho) =
	(1 - p) \rho \otimes \dyad{0}
	+ p \frac{I}{2} \otimes \dyad{1},
\end{equation}
erases a qubit with probability $p$ by replacing it with a maximally mixed single-qubit state.
The second register indicates whether a qubit has been erased.
Since
\begin{equation}
	\frac{I}{2} = \frac{1}{4}\qty(I\rho I + X \rho X + Y \rho Y + Z \rho Z),
\end{equation}
the multi-qubit version of the channel can be written as
\begin{equation}
	\mathcal{E}^{n}_p(\rho) =
	\sum_{\vb e \in \qty{0, 1}^n}
	\Pr[\vb e]
	\qty(
	\frac{1}{4^{|\vb e|}}
	\sum_{E \in P_n(\vb e)} E\rho E
	)
	\otimes \dyad{\vb e},
\end{equation}
where $P_n(\vb e)$ is the set of $n$-qubit Pauli operators
with a trivial action on every qubit $q_i$ for which  $e_i = 0$.
Then,
after a measurement of the second register identifying erasure positions,
the channel reduces to a Pauli channel on the erased qubits
where each error is equally likely.
Therefore,
from a measurement of the syndrome,
a maximum likelihood decoder searches for any Pauli operator restricted
to the erased region with the appropriate syndrome.

The success probability of this decoder is the inverse of the number
of logical operators that cannot be moved out of the region of erased qubits by
multiplying with a stabilizer.
This probability can be computed by Gaussian reduction
as described in~\cite{gullans_quantum_2021}.

\begin{figure}
	\begin{center}
		\includegraphics{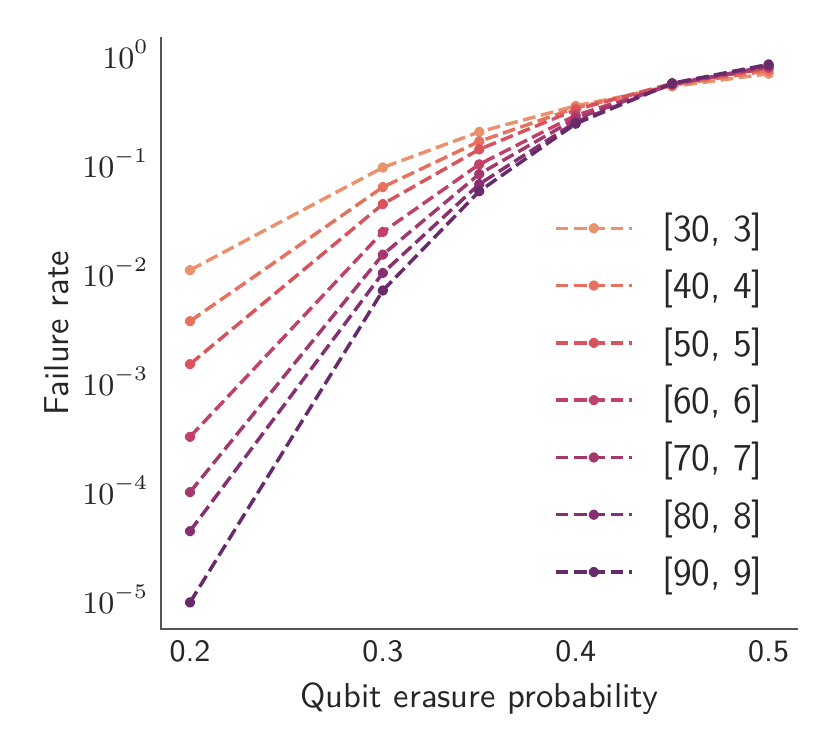}
	\end{center}
	\caption{
		Failure rate for maximum-likelihood decoding of the erasure channel.
		For each system size, we show the lowest failure rates amongst all codes found
		for $\delta_q = 3$, $\delta_s = 6$ and $\Delta_s = 20$.
		The codes used in this experiment are available online~\cite{tremblay_maxime_2023_7658784}.
	}
	\label{fig:threshold}
\end{figure}

The capacity, i.e. the maximum rate of information,
of the erasure channel~\cite{bennett_capacities_1997} is computed from the probability of erasure.
That is,
the capacity of the channel with erasure probability $p$ is $R_{\max} = 1 - 2p$.
Inverting this relation,
we observe that a code family with rate $R$ can be used to protect information
as long as the erasure probability is at most $\frac{1 - R}{2}$.
Figure~\ref{fig:threshold} illustrates that the rate $1/10$ codes obtained with our methodology achieve this limit.
In other words,
we have a procedure to construct random sparse stabilizer codes that are essentially
capacity-achieving for the erasure channel.

\section{Discussion and outlook}
\label{sec:conclusion}

In this work, we reformulate the search for finite-rate sparse stabilizer codes as a constraint satisfaction problem, which we call stabilizing edge coloring, that is amenable to solution with state-of-the-art CSP solvers.
We believe future work can exploit the connection between quantum error correction and constraint programming that we establish here in order to guide the search for new
families of random stabilizer codes.

We note that even though here we limit our analysis to CSS codes for the sake of simplicity,
the method we introduce is easily adaptable to more general stabilizer codes.
For example,
one can assign an extra variable to each edge to represent its Pauli value
instead of labelling the stabilizer vertices directly.
Furthermore, the method is flexible in the sense that it is simple
to incorporate constraints that take into account multiple factors on the same footing,
including, for example, hardware limitations like qubit layout and connectivity.

In this work we had to strike a balance between computational resources and a reasonable timeframe for completion of numerical calculations.
Targeted searches for larger codes within the satisfiable phase using more resources are a straightforward direction for future work.
Impressive progress in the performance of solvers in the last decades~\cite{noauthor_sat_nodate, noauthor_minizinc_nodate} means that larger codes could soon be discoverable by our methodology with moderate resources.
Customized CSP solvers that resolve the types of constraints involved in stabilizing edge coloring could also boost the search for codes with desirable properties.

It is also interesting to study different random graph constructions as
input to CSP solvers instead of the uniformly sampled graphs we used.
This could lead to sparser initial graphs with structure favorable to the discovery of good quantum codes.

Finally, we were able to use our approach to construct a family of finite-rate codes
achieving optimal threshold for the erasure channel.
This result, together with many recent stabilizer code constructions,
motivate the search for more generally applicable quantum decoders for
more complex noise channels such as the depolarizing channel and correlated Pauli noise.
This would allow us to probe the performance of random code constructions more
thoroughly.

\section*{Acknowledgments}

This work was supported by the Ministère de l'Économie et de l'Innovation du Québec via its Research Chair in Quantum Computing and a Natural Sciences and Engineering Research Council of Canada Discovery grant. MT is supported by a Canada Graduate Scholarship from the Natural Sciences and Engineering Research Council of Canada. We acknowledge Calcul Québec and Compute Canada for computing resources.
We thank the members of the QuICoPhy Theory Lab for valuable discussions.

\bibliographystyle{plainnat}
\bibliography{references}

\onecolumn\newpage
\appendix

\section{Exploration of different degree bounds}
\label{sec:extra}

\begin{figure}[t]
	\begin{center}
		\begin{subfigure}[b]{0.3\textwidth}
			\centering
			\includegraphics[scale=0.7]{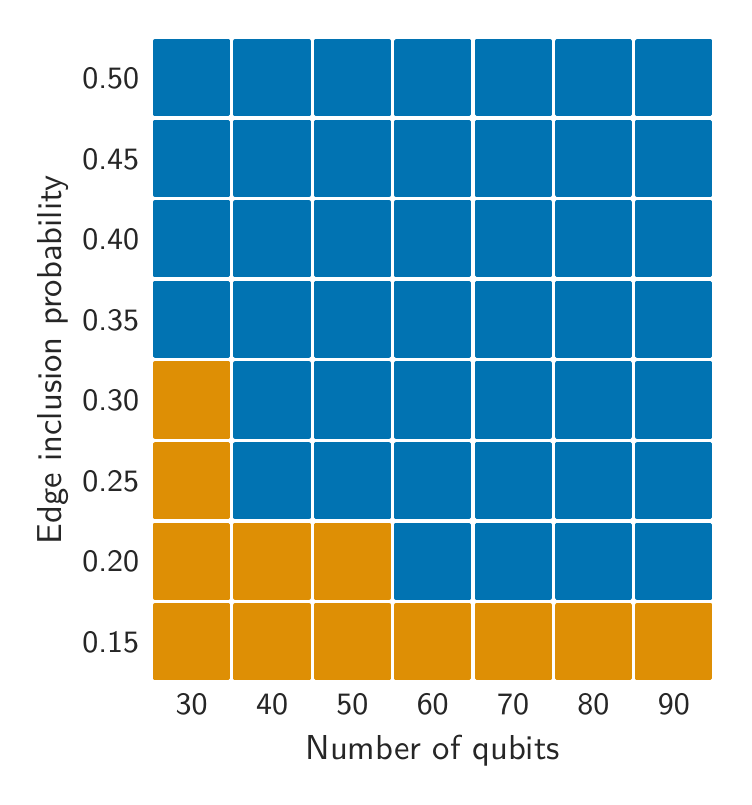}
			\caption{
				$\delta_q = 2$, $\delta_s = 4$, $\Delta_s = 15$
			}
			\label{fig:sat_phase_2_15}
		\end{subfigure}
		\begin{subfigure}[b]{0.3\textwidth}
			\centering
			\includegraphics[scale=0.7]{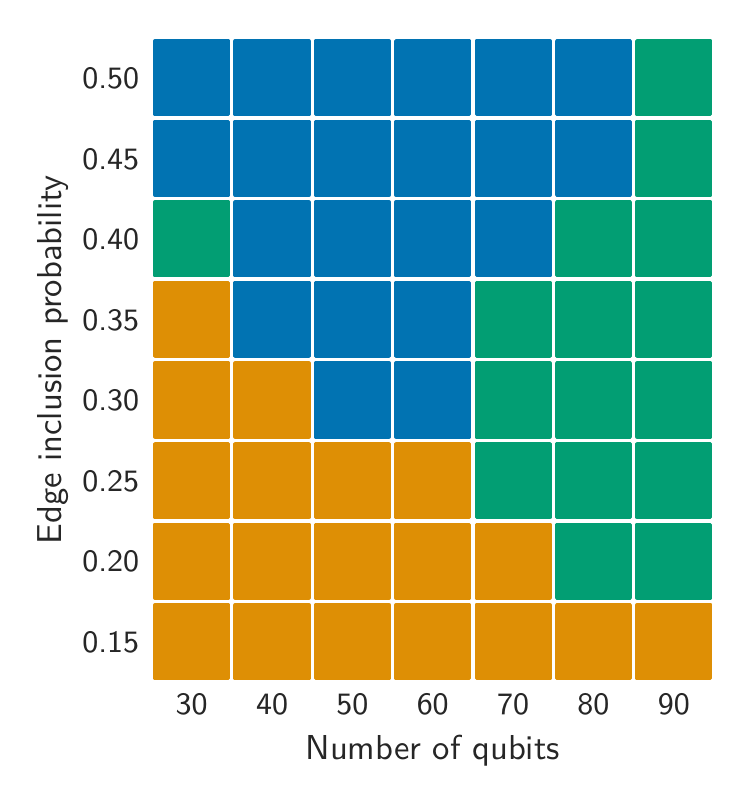}
			\caption{
				$\delta_q = 3$, $\delta_s = 6$, $\Delta_s = 15$
			}
			\label{fig:sat_phase_3_15}
		\end{subfigure}
		\begin{subfigure}[b]{0.3\textwidth}
			\centering
			\includegraphics[scale=0.7]{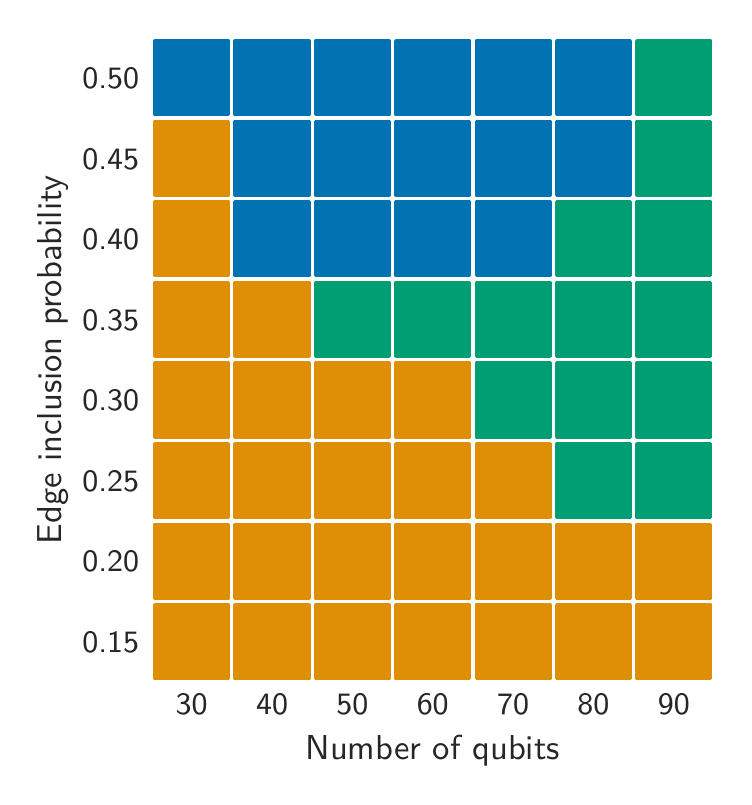}
			\caption{
				$\delta_q = 4$, $\delta_s = 8$, $\Delta_s = 20$
			}
			\label{fig:sat_phase_4_20}
		\end{subfigure}
	\end{center}
	\caption{
		Satisfiability phase diagrams with balancing contraints and different degree bounds.
		Each pixel is generated by sampling 10 support graphs and
		all parameters and details are the same as in Figure~\ref{fig:sat_phase_qubit_deg}.
	}
	\label{fig:sat_phase_other}
\end{figure}

In the main text,
we argue that other choices of degree bounds lead to qualitively similar results.
To support that argument,
we show in Figure~\ref{fig:sat_phase_other} satisfiability phase diagrams
for three other set of degree bounds.
Each diagram is generated using the same hardware and time limit
as those of Figure~\ref{fig:sat_phase_qubit_deg} and Figure~\ref{fig:sat_phase}.

In all three scenarios,
we observe similar transition between the satisfiable and unsatisfiable phases.
Due to computational restrictions,
these diagrams are generated using ten samples per pixel.
This induces a relatively large unknown region in Figures~\ref{fig:sat_phase_3_15}
and~\ref{fig:sat_phase_4_20} which are both obtained from CSP instances strictly harder than the
one presented in the main text.
However,
more samples and a longer time limit,
would narrow down both unknown phases to a smaller area.
This would require significantly more computational resources and is left to further investigation.

\end{document}